\documentclass[12pt]{article}
\usepackage{epsfig,amsmath,amssymb}
\usepackage{psfrag}

   {\end{list}}%

\def\a{\alpha}
\def\b{\beta}

\def\m{\mu}
\def\n{\nu}
\def\r{\rho}
\def\s{\sigma}

\def\prslash{{\partial\mkern-9mu/}}
\def\pslash{{p\mkern-8mu/}{\!}}

\def\prslash{{\partial\mkern-9mu/}}    
\def\qslash{{q\mkern-8mu/}{\!}}

\def\iDq{\int\!\! \frac{d^D\!q}{(2\pi)^D}}

\def\idx{\int\!\! d^4\!x}

\newcommand{\bea}{\begin{eqnarray}}
\newcommand{\eea}{\end{eqnarray}}
\newcommand{\beann}{\begin{eqnarray*}}
\newcommand{\eeann}{\end{eqnarray*}}
\newcommand{\ba}{\begin{array}}
\newcommand{\ea}{\end{array}}

\newcommand{\Tr}{\mathbf{Tr}}

\def\psib{\bar{\psi}}
\def\Psib{\bar{\Psi}}
\def\g5{\gamma_{5}}

\def\prslash {{\partial\mkern-9mu/}}  
\def\pslash  {{p\mkern-7mu/}}

\def\idx{\int\! d^{4}\!x\,}

 \def\psib{\bar{\psi}}
 \def\Psib{\bar{\Psi}}

 \def\prslash {{\partial\mkern-9mu/}}  
 \def\pslash  {{p\mkern-7mu/}}
\def\qslash  {{q\mkern-7mu/}}

 \def\Dm {{\partial}_{\mu}}
 \def\Dn {{\partial}_{\nu}}

 \def\g {\gamma}

 \def\a {\alpha}
\def\b {\beta}
\def\r {\rho}
 \def\s {\sigma}

 \def\Tr{\text{Tr}}
\def\tr{\text{tr}}

\def\id{{\rm{I}\!\rm{I}}}

\psfrag{al}{\scriptsize$\alpha$}
\psfrag{be}{\scriptsize$\beta$}
\psfrag{mu}{\scriptsize$\mu$}
\psfrag{nu}{\scriptsize$\nu$}
\psfrag{Psib}{\scriptsize$\Psib$}
\psfrag{Psi}{\scriptsize$\Psi$}
\psfrag{Phic}{\scriptsize$\Phi^*$}
\psfrag{Phi}{\scriptsize$\Phi$}
\psfrag{p}{\scriptsize p}
\psfrag{q}{\scriptsize q}
\psfrag{r}{\scriptsize r}
\psfrag{s}{\scriptsize s}
\psfrag{I}{\scriptsize I}
\psfrag{J}{\scriptsize J}
\psfrag{K}{\scriptsize K}
\psfrag{L}{\scriptsize L}
\psfrag{M}{\scriptsize M}
\psfrag{N}{\scriptsize N}

\begin{document}
\begin{titlepage}
\rightline{FTI/UCM 75-2007}
\vglue 45pt

\begin{center}

{\Large \bf Renormalisability of the matter  determinants in
noncommutative gauge theory in the enveloping-algebra formalism}\\
\vskip 1.2 true cm
{\rm C. P. Mart\'{\i}n}\footnote{E-mail: carmelo@elbereth.fis.ucm.es}
 and C. Tamarit\footnote{E-mail: ctamarit@fis.ucm.es}
\vskip 0.3 true cm
{\it Departamento de F\'{\i}sica Te\'orica I,
Facultad de Ciencias F\'{\i}sicas\\
Universidad Complutense de Madrid,
 28040 Madrid, Spain}\\
\vskip 0.75 true cm

\vskip 0.25 true cm

{\leftskip=50pt \rightskip=50pt
\noindent
We consider noncommutative gauge theory defined by means of Seiberg-Witten maps for an arbitrary semisimple gauge group. We compute the one-loop UV divergent matter contributions to the gauge field effective action to all orders in the noncommutative parameters $\theta$. We do this for Dirac fermions  
and complex scalars carrying arbitrary representations of  
the gauge group. We use path-integral methods in the framework of dimensional regularisation and consider arbitrary invertible Seiberg-Witten maps that are linear in the matter fields. Surprisingly, it turns out that the UV divergent parts of the matter contributions are proportional to the noncommutative Yang-Mills action where traces are taken over the representation of the matter fields; this result supports the need to include such traces in 
the classical action of the gauge sector of the noncommutative theory.

\par }
\end{center}

\vspace{20pt}
\vfill
\end{titlepage}


\setcounter{page}{2}

	The issue of renormalisability of noncommutative gauge theories in the enveloping-algebra approach has been a subject of 
 intense research in the last years \cite{Bichl:2001cq,Wulkenhaar:2001sq,Buric:2002gm,Buric:2004ms,Buric:2005xe,Buric:2006wm,Calmet:2006zy,Martin:2006gw,Latas:2007eu}. The outcome of this research so far shows that NC Yang Mills is one-loop renormalisable up to first order in $\theta$ \cite{Buric:2005xe,Latas:2007eu} ---in fact, up to order two for the case of NC U(1) Yang-Mills \cite{Buric:2002gm}--- but renormalisability is spoiled by the presence of Dirac fermions in the fundamental repesentation or complex scalars in the $U(1)$ case \cite{Wulkenhaar:2001sq,Buric:2002gm,Buric:2004ms,Martin:2006gw}. However, in all cases the gauge sector of the theory remains renormalisable despite the presence of matter; this has also been checked for a noncommutative extension of the Standard Model \cite{Buric:2006wm} in which the traces in the gauge sector are taken over all the different particle representations. This renormalisability of the gauge sector is quite intriguing and
far from trivial since BRS invariance and power-counting do not account for it.
Indeed, take a simple compact  gauge group, then, power-counting and BRS invariance do not restrict the one-loop UV divergent part of the effective action of the gauge field in the background-field gauge to the noncommutative Yang-Mills action, but to a linear combination with arbitrary UV divergent coefficients of the noncommutative Yang-Mills action and terms like 
$\theta^{\alpha\beta}\text{Tr}\idx \,F_{\alpha\beta}\star 
F_{\m\n}\star F^{\m\n}$, $\theta^{\alpha\beta}\text{Tr}\idx \,F_{\mu\alpha}\star F_{\beta\n}\star F^{\m\n}$, etc ... 
   The confirmation of the renormalisablity we have mentioned at higher orders in $\theta$ and its understanding --perhaps, as a by-product of an as yet 
undiscovered symmetry of the theory--, as well as the study of its dependence on the choice of traces for the noncommutative Yang-Mills action, are still open problems.

 As a first step in this direction, in this paper, we compute to all orders in $\theta$ the UV part of the 
one-loop effective action obtained by integrating out the matter fields in noncommutative gauge theory for arbitrary semisimple gauge group.
By ``matter'' we mean Dirac fermions and complex scalars in an arbitrary unitary irreducible representation of the gauge group. What we have obtained is that, 
in both cases and in dimensional regularisation with $D=4+2\epsilon$, the pole part  
of the effective action of the gauge field turns out to be  proportional to 
the noncommutative Yang-Mills action with traces taken over the representation of the gauge group acting on the matter fields, namely:
\begin{equation}
\begin{array}{l}
 \Gamma^{f}[A]^\text{one-loop}_\text{pole,$A$-dep.}=\frac{1}{48\pi^2\epsilon}\,{\textbf{\Tr}}\idx \,F_{\m\n}\star F^{\m\n},\\
 \Gamma^{sc}[A]^\text{one-loop}_\text{pole,$A$-dep.}=\frac{1}{192\pi^2\epsilon}\,{\textbf{\Tr}}\idx\,F_{\m\n}\star F^{\m\n},\,\,F_{\m\n}=\Dm A_\n-\Dn A_\m-i[A_\m,A_\n]_\star.
\end{array}
\label{mainresult}
\end{equation}
This result supports the need to consider such types of traces for models aiming to have the renormalisability property; in fact, all models considered so far with a one-loop, order-$\theta$ renormalisable gauge sector have these types of traces. A relevant example is the noncommutative version of the Standard Model in ref.~\cite{Buric:2006wm}, whose gauge sector involves a non-trivial sum of traces over all the particle representations of the model. 
Our result holds for the class of Seiberg-Witten applications for which the map between the noncommutative and ordinary matter fields is linear and invertible; the computation to all orders in $\theta$ is feasible due to the possibility of changing variables in the functional integrals from the ordinary fields to the noncommutative fields. The result we have obtained is quite surprising since BRS invariance and power-counting of the theory formulated in terms of the ordinary fields do not enforce it, and, it is relevant in the phenomenological applications of noncommutative gauge theories, since it supports the robustness of the predictions based on the gauge sector of the theory. These phenomenological predictions can certainly be tested at the LHC~\cite{Buric:2007qx,Alboteanu:2006hh,MohammadiNajafabadi:2006iu,Behr:2002wx}.

	Using a similar notation as that employed in ref.~\cite{Brandt:2003fx}, we  consider a noncommutative gauge theory 
with a semisimple gauge group of the form $G_1\times\dots\times G_N$ with $G_i$ simple for $i=1\dots s$ and abelian for $i=s+1,\dots,N$. Then the ordinary gauge field will be of the form
$$
a_\m=\sum_{k=1}^sg_k(a^k_\m)^a(T^k)^a+\sum_{l=s+1}^Ng_l a_\m^lT^l,$$
where the $T's$ are generators of unitary irreducible representations of the group factors. The matter fields to consider are Dirac fermions $\psi$ and complex scalars $\phi$ in an irreducible representation of $G$ and therefore carrying multi-indices $I=i_1\dots i_s$ for the irreducible factors. In multi-index notation we can define generators and a ``global'' trace $\bf{Tr}$ as follows
\begin{equation*}
\begin{array}{l}
(T^k)^a_{IJ}=\delta_{i_1 j_1}\cdots(T^k)^a_{i_k j_k}\cdots \delta_{i_s j_s},\quad k=1,\dots, s,\\
T^l_{IJ}= \delta_{i_1 j_1}\cdots \delta_{i_s j_s}Y^l,\quad l=s+1,\dots,N,\\
\text{\bf\Tr}(T^k)^a(T^{k'})^{a'}=(T^k)^a_{IJ}(T^{k'})^{a'}_{JI}.
\end{array}
\end{equation*}
	In order to build noncommutative actions for the matter fields we need the Seiberg-Witten maps for the noncommutative 
gauge field $A_\m$ and the matter fields, i.e., noncommutative fermions $\Psi_{\a I}$ and complex scalars $\Phi_I$. We make no assumption on the map for the gauge field, but for the matter fields we consider maps of the form
\begin{equation}
  \Psi_{\a I}=(\delta_{IJ}\delta_{\a\b}+M[a_\m,\partial,\gamma;\theta]_{\a\b I J})\psi_{\b J},\quad\quad
 \Phi_I=(\delta_{IJ}+N[a_\m,\partial;\theta]_{I J})\phi_{J},
\label{SW}
\end{equation}
and analogously for the Dirac adjoint fermion $\Psib=\Psib^\dagger\g^0$ and the complex conjugate $\Phi^*$ of $\Phi$. $\g^\m$ denotes the gamma matrices satisfying $\{\g^\m,\g^\n\}=2\eta^{\m\n}$.

	With this notation we define next the actions for the matter fields in noncommutative spacetime
\begin{equation}
\begin{array}{l}
S^\text{f}=\idx(\Psib\star i\g^\m D_{\m}\Psi-m\Psib\star\Psi),\quad   
S^\text{sc}=\idx ((D_\m \Phi)^*\star D^\m\Phi-m^2\Phi^*\Phi)-V(\Phi),\\
(D_\m)_{IJ}=\delta_{IJ}\partial_\m-i A_{\m\,IJ}\star,
\end{array}
\label{S}
\end{equation}
where $\star$ denotes the usual Moyal product and $V(\Phi)$ is an arbitrary noncommutative gauge-invariant potential that will not contribute to the $A_\m$-dependent part of the gauge effective action. Our objective is to compute the divergent part of the one-loop  effective actions
for the gauge field that are formally defined by
\begin{equation}
\begin{array}{l}
 \Gamma^{f/sc}[a;\theta]=-i\ln Z^{f/sc}[a;\theta],\\
Z^{f}[a;\theta]={\cal N}^f\int [d\psib][d\psi]\exp(iS^f),
\quad Z^{sc}[a;\theta]={\cal N}^{sc}\int [d\phi^*][d\phi]\exp(iS^{sc}),\\
{\cal N}^{f/sc}=Z^{f/sc}[0;\theta]^{-1}.
\end{array}
\label{Gamma}
\end{equation}
The previous expressions relate the gauge effective actions to the determinants of the operators appearing in the actions for the matter fields. We will make sense out of these formal definitions by using dimensional regularisation in $D=4+2\epsilon$ dimensions; this will make the divergent contributions appear as poles in $\epsilon$. Note that, in order to work out the UV divergence of $\Gamma^{f/sc}$ to all orders in $\theta$ by integrating out the matter fields in the functional integrals in eq.~\eqref{Gamma}, one would need to know the Seiberg-Witten map to all these orders. We can avoid this by
performing a change of variables in the functional integrals from the ordinary fields to the noncommutative ones. Recalling eq.~\eqref{SW},
\begin{equation*}
 \begin{array}{l}
  [d\psib][d\psi]=\text{det}(\id+M)\text{det}(\id+\bar{M})[d\Psib][d\Psi],\\
 \phantom{}[d\phi^*][d\phi]=\text{det}(\id+N)^{-1}\text{det}(\id+N^*)^{-1}[d\Phi^*][d\Phi].
 \end{array}
\end{equation*}
	The above determinants are defined in dimensional regularisation by a diagrammatic expansion where the propagators
are equal to the identity. As a consequence they are given in momentum space by tadpole-like integrals, which are zero. Therefore,
\begin{equation}
 Z^{f}[a;\theta]={\cal N}^f\int [d\Psib][d\Psi]\exp(iS^f),\quad Z^{sc}[a;\theta]={\cal N}^{sc}\int[d\Phi^*][d\Phi]\exp(iS^{sc})
\label{Z}
\end{equation}
and, since the matter actions given in eq.~\eqref{S} depend on the noncommutative gauge field $A_\m$, we have that the dependence of the effective actions $\Gamma$ on $a_\m$ is through a dependence on $A_\m$:
$$\Gamma^{f/sc}[a;\theta]=\Gamma^{f/sc}[A;\theta].$$
In particular, when the eqs.~\eqref{Gamma} and \eqref{Z} defining $\Gamma^{f/sc}$ are interpreted diagramatically, it is clear that the potential $V(\Phi)$ makes no contribution to the $A_\m$-dependent part of the gauge effective actions. Thus,
eqs.~\eqref{Gamma} and \eqref{Z} allow us to obtain the $A_\m$-dependent parts of the effective actions $\Gamma^{f/sc}$ as the determinants of the operators
that appear in the actions for the matter in terms of the noncommutative 
fields. Integrating over $[d\Psib],\,[d\Psi]$ and
$[d \Phi^*],\,[d\Phi]$ neglecting $V(\Phi)$ we obtain:
\begin{align}
 \nonumber i\Gamma^f[A]_{A\text{-dep.}}&\!=\!\ln\frac{\text{det}[\prslash+im-iA\mkern-9mu/\star]}{\text{det}[\prslash+im]}=\Tr\ln[1-(\prslash+im)^{-1}iA\mkern-9mu/\star]=-\!\sum_{n=1}^\infty \frac{1}{n}\,\Tr\,[(\prslash+im)^{-1}iA\mkern-9mu/\star]^n,\\
 i\Gamma^{sc}[A]_{A\text{-dep.}}&\!\!=\!-\!\ln\!\frac{\text{det}[iD^2\!+\!im^2]}{\text{det}[i\partial^{\,2}\!+\!im^2]}\!=\!\!\sum_{n=1}^\infty\frac{(\!-\!1)^n}{n}\Tr[(i\partial^{\,2}\!+\!i\!m^2)^{-1}((\partial\!\cdot A)\star+2A\cdot\partial\star-iA_\m\star A^\m \star))]^n.
\label{Gammas}
\end{align}
In the previous expressions  $\Tr$ denotes a trace over discrete indices and integration over the continuous indices of the corresponding operators. The operators $(\prslash+im)^{-1}$ and $(i\partial^{\,2}\!+i\!m^2)^{-1}$ have matrix elements given by ordinary propagators:
\begin{equation}
\langle y| (\prslash+im)^{-1}|x\rangle=\int\!\frac{d^Dp}{(2\pi)^D}\,\frac{ie^{-ip(y-x)}(\pslash+m)}{p^2-m^2+i0^+},\,\langle y| (i\partial^{\,2}\!+\!im^2)^{-1}|x\rangle=\int\!\frac{d^Dp}{(2\pi)^D}\,\frac{ie^{-ip(y-x)}}{p^2-m^2+i0^+}.
\label{propagators}
\end{equation}
 Since we are interested in the divergent part of $\Gamma^{f/sc}$, we need to identify the contributions in eq.~\eqref{Gammas} that yield the poles at $D=4$. 
This can be done by using power-counting arguments as follows. Let us first 
note that the propagators in eq.~\eqref{propagators} are diagonal in the colour indices, then, one realises that the trace in eq.~\eqref{Gammas} forces a trace in the colour indices of the background gauge fields $A_{\m IJ}$. Therefore we can write
\begin{equation}
i\Gamma^{f/sc}[A]_{A\text{-dep.}}\equiv\sum_{n=1}^\infty\int\!d^dx_1\dots\int\!d^dx_n\textbf{\Tr}[A_{\mu_1}(x_1)\dots A_{\mu_n}(x_n)]
\Gamma^{f/sc (n)\,\m_1\dots\m_n}[x_1,\dots,x_n],
\label{Gamman}
\end{equation}
where the mass dimension of $\Gamma^{f/sc (n)\,\m_1\dots\m_n}[x_i]$ is, for $D=4$, 4+3n. In momentum space $\Gamma^{f/sc (n)\,\m_1\dots\m_n}[p_i]$ will be of dimension $4-n$ and, due to the translation invariance of the propagators in eq.~\eqref{propagators}, it is given by a single loop integral. Then, power-counting
tell us that $\Gamma^{f/sc (n)}$ with $n\geq5$ are finite. Thus we 
only need to work out contributions with up to 4 background gauge 
fields $A_\m$.
	Let us start with the case of fermions. Going over to momentum space and starting from eq.~\eqref{Gammas}, we can express 
$\Gamma^{f(n)}[A]$ in terms of a loop integral as follows
\begin{align}
 \nonumber&\Gamma^{f(n)}_{\m_1\dots\m_n}[x_1,\dots,x_n]=\frac{(-1)^{n+1}}{n}\int\prod_{i=1}^n\frac{d^Dp_i}{(2\pi)^D}(2\pi)^D\delta\big(\sum_i p_i\big)e^{i\sum_{i=1}^np_i x_i}e^{-i\sum_{i<j}^n p_i\circ p_j}\times\\
&\times\int\!\frac{dq}{(2\pi)^D}\,\frac{\tr[(\qslash+\pslash_1+m)\g^{\m_1}(\qslash+m)\g^{\m_2}(\qslash-\pslash_2+m)\cdots(\qslash-\sum_{i=2}^{n-1}\pslash_i+m)\g^{\m_n}]}{[(q+p_1)-m^2][q^2-m^2][(q-p_2)^2-m^2]\cdots[(q-\sum_{i=2}^{n-1}p_i)^2-m^2]}.
\label{Gammanf}
\end{align}
	$p\circ q\equiv \frac{i}{2}\theta^{\a\b}p_{\a}q_\b$.
	In the case of scalars, from the formula in eq.~\eqref{Gammas} it is clear that the expansion in $n$ does not correspond
to an expansion in the number of background fields due to the presence of interaction vertices with one and two $A_\m's$. Instead of a closed formula for $\Gamma^{(n)}$ as the one just given for fermions in eq.~\eqref{Gammanf}, we provide formulae for the potentially divergent contributions $\Gamma^{sc(k)},\,1\leq k\leq4$.
\begin{align}
 \nonumber&\Gamma^{sc(n)}_{\m_1\dots\m_n}[x_1,\dots,x_n]=\int\prod_{i=1}^n\frac{d^Dp_i}{(2\pi)^D}(2\pi)^D\delta\big(\sum_i p_i\big)e^{i\sum_{i=1}^np_i x_i}\tilde{\Gamma}^{(n)\m_1\dots\m_n}[p_1,\dots,p_n],\\
\nonumber&\tilde{\Gamma}^{(1)\m}[0]=-\iDq\,\frac{2q^\m}{q^2-m^2},\\
\nonumber&\tilde{\Gamma}^{(2)\m\n}[p_1,p_2]=-\iDq\,\frac{e^{-p_1\circ p_2}\eta^{\m\n}}{q^2-m^2}+\frac{1}{2}\iDq\,\frac{e^{-p_1\circ p_2}(2q+p_1)^{\m}(2q+p_1)^{\n}}{[(q+p_1)^2-m^2]^2},\\
\nonumber&\tilde{\Gamma}^{(3)\m\n\r}[p_1,p_2,p_3]=\iDq\,\frac{e^{-\sum_{i<j}p_i\circ p_j}\eta^{\n\r}(2q-p_1)^\m}{[q^2-m^2][(q-p_1)^2-m^2]}\\
\label{Gammansc}&\phantom{\tilde{\Gamma}^{(3)\m\n\r}[p_1,p_2,p_3]}-\frac{1}{3}\iDq\,\frac{e^{-\sum_{i<j}p_i\circ p_j}(2q+p_1)^{\m}(2q-p_2)^{\n}(2q-p_2+p_1)^\r}{[q^2-m^2][(q-p_2)^2-m^2][(q+p_1)^2-m^2]},\\
\nonumber&\tilde{\Gamma}^{(4)\m\n\r\s}[p_1,p_2,p_3,p_4]=\frac{1}{2}\iDq\,\frac{e^{-\sum_{i<j}p_i\circ p_j}\eta^{\m\n}\eta^{\r\s}}{[q^2-m^2][(q-p_3-p_4)^2-m^2]}\\
\nonumber&\phantom{\tilde{\Gamma}^{(3)\m\n\r}[p_1,p_2,p_3]}-\iDq\,\frac{e^{-\sum_{i<j}p_i\circ p_j}\eta^{\m\n
}(2q-p_3)^{\r}(2q-2p_3-p_4)^{\s}}{[q^2-m^2][(q-p_3)^2-m^2][(q-p_3-p_4)^2-m^2]}\\
\nonumber&\phantom{\tilde{\Gamma}^{(3)\m\n\r}[p_1]}+\frac{1}{4}\iDq\,\frac{e^{-\sum_{i<j}p_i\circ p_j}(2q+p_1)^{\m}(2q-p_2)^{\n}(2q-2p_2-p_3)^\r(2q-p_2-p_3+p_1)^\s}{[q^2-m^2][(q-p_2)^2-m^2][(q-p_2-p_3)^2-m^2][(q+p_1)^2-m^2]}.
\end{align}
 Note that the phases in the expressions in eqs.~\eqref{Gammanf} and \eqref{Gammansc} are independent of the loop momentum; this makes it possible to obtain the corresponding UV divergent contribution at all orders in $\theta$. The final expressions can be obtained from $\Gamma^{f/sc(n)}[x_i]$ for $1\leq n\leq 4$ by using eqs.~\eqref{Gammansc}, \eqref{Gammanf}  and \eqref{Gamman} and have been given in eq.~\eqref{mainresult} already.
The results in eq.~\eqref{mainresult} hold independently of the form of scalar potential $V(\Phi)$. In particular, they remain valid in the case of spontaneous symmetry breaking.
	The previous expressions are proportional to possible terms in the gauge sector piece of the noncommutative Yang-Mills 
lagrangian  involving the traces $\textbf{Tr}$ over the matter representations, and the UV divergencies can be substracted by the ordinary renormalisation of the  couplings corresponding to such terms. 

Finally, we believe that the results presented in this paper will also hold for noncommutative gauge theories with chiral fermions, 
if they are anomaly free. Unfortunately, even in the ordinary case --see ref.~\cite{Martin:1999cc}-- a rigorous derivation of the results analogous to 
those presented in this paper, will demand that one carries out far more involved computations and also a separate paper.



\begin{thebibliography}{99}

\bibitem{Bichl:2001cq}
  A.~Bichl, J.~Grimstrup, H.~Grosse, L.~Popp, M.~Schweda and R.~Wulkenhaar,
  JHEP {\bf 0106} (2001) 013
  [arXiv:hep-th/0104097].


\bibitem{Wulkenhaar:2001sq}
  R.~Wulkenhaar,
  JHEP {\bf 0203} (2002) 024
  [arXiv:hep-th/0112248].


\bibitem{Buric:2002gm}
  M.~Buric and V.~Radovanovic,
  JHEP {\bf 0210} (2002) 074
  [arXiv:hep-th/0208204].

\bibitem{Buric:2004ms}
  M.~Buric and V.~Radovanovic,
  JHEP {\bf 0402} (2004) 040
  [arXiv:hep-th/0401103].

\bibitem{Buric:2005xe}
  M.~Buric, D.~Latas and V.~Radovanovic,
  JHEP {\bf 0602} (2006) 046
  [arXiv:hep-th/0510133].

\bibitem{Buric:2006wm}
  M.~Buric, V.~Radovanovic and J.~Trampetic,
  JHEP {\bf 0703}, 030 (2007)
  [arXiv:hep-th/0609073].

\bibitem{Calmet:2006zy}
  X.~Calmet,
  Eur.\ Phys.\ J.\  C {\bf 50} (2007) 113
  [arXiv:hep-th/0604030].




\bibitem{Martin:2006gw}
     C.~P.~Martin, D.~Sanchez-Ruiz and C.~Tamarit,
   JHEP {\bf 02} (2007) 065 
  [arXiv:hep-th/0612188].

\bibitem{Latas:2007eu}
  D.~Latas, V.~Radovanovic and J.~Trampetic,
  arXiv:hep-th/0703018.




\bibitem{Buric:2007qx}
  M.~Buric, D.~Latas, V.~Radovanovic and J.~Trampetic,
  Phys.\ Rev.\  D {\bf 75} (2007) 097701.

\bibitem{Alboteanu:2006hh}
  A.~Alboteanu, T.~Ohl and R.~Ruckl,
  Phys.\ Rev.\  D {\bf 74} (2006) 096004
  [arXiv:hep-ph/0608155].

\bibitem{MohammadiNajafabadi:2006iu}
  M.~Mohammadi Najafabadi,
  Phys.\ Rev.\  D {\bf 74} (2006) 025021
  [arXiv:hep-ph/0606017].

\bibitem{Behr:2002wx}
  W.~Behr, N.~G.~Deshpande, G.~Duplancic, P.~Schupp, J.~Trampetic and J.~Wess,
  Eur.\ Phys.\ J.\  C {\bf 29} (2003) 441
  [arXiv:hep-ph/0202121].

\bibitem{Brandt:2003fx}
  F.~Brandt, C.~P.~Martin and F.~R.~Ruiz,
  JHEP {\bf 0307} (2003) 068
  [arXiv:hep-th/0307292].





\bibitem{Martin:1999cc}
  C.~P.~Martin and D.~Sanchez-Ruiz,
  Nucl.\ Phys.\  B {\bf 572} (2000) 387
  [arXiv:hep-th/9905076].

\end{thebibliography}
\end{document}